\documentclass[prl,twocolumn,amsmath,amssymb,showpacs]{revtex4}

\usepackage{citesort}

\usepackage{graphicx}

\begin{document}

\title{Fluctuation-Driven Molecular Transport in an Asymmetric Membrane Channel}

\author{Ioan Kosztin}
\affiliation{Department of Physics \& Astronomy,
  University of Missouri, Columbia, MO 65211, USA} 
\author{Klaus Schulten}
\affiliation{Beckman Institute and Department of Physics, University of
  Illinois, Urbana, IL 61801, USA}

\date{August 20, 2004}

\begin{abstract}
  Channel proteins, that selectively conduct molecules across cell
  membranes, often exhibit an asymmetric structure.  By means of a
  stochastic model, we argue that channel asymmetry in the presence of
  non-equilibrium fluctuations, fueled by the cell's metabolism as
  observed recently, can dramatically influence the transport through
  such channels by a ratchet-like mechanism.  For an aquaglyceroporin
  that conducts water and glycerol we show that a previously determined
  asymmetric glycerol potential leads to enhanced inward transport of
  glycerol, but for unfavorably high glycerol concentrations also to
  enhanced outward transport that protects a cell against poisoning.
\end{abstract}


\maketitle

Living cells interact with their extracellular environment through the cell
membrane, which acts as a protective permeability barrier for preserving the
internal integrity of the cell. However, cell metabolism requires controlled
molecular transport across the cell membrane, a function that is fulfilled by
a wide variety of transmembrane proteins, acting as passive and active
transporters \cite{ALBE2002}. Channel proteins as passive transporters
facilitate the diffusion of specific molecules across the membrane down a free
energy gradient.
Active transporters conduct molecules along or against the free energy
gradient consuming for that purpose external energy.
However, the plasma membrane of living cells is subject to a variety of
non-equilibrium, i.e., non-thermal processes, e.g., interaction with the
cytoskeleton and with active membrane proteins
\cite{levin91-733,ramaswamy00-3494}.
In this Letter we want to argue that in an active plasma membrane even passive
channel proteins can act as active transporters by consuming energy from
non-equilibrium fluctuations fueled by cell metabolism
\cite{pelling04-1147}.

\begin{figure}[!htp]
\includegraphics[width=3.2in]{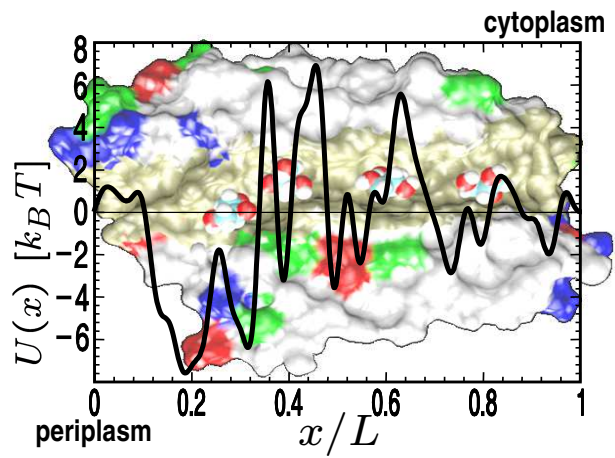}
\caption{(color online). Section through the glycerol conduction pathway in GlpF and
  the corresponding asymmetric PMF (solid curve) reported in
  \cite{jensen02-6731}.  }
\label{fig:pmf}
\end{figure}

The \emph{Escherichia coli} glycerol uptake facilitator (GlpF) is an
aquaglycerol channel protein, which transports both water and glycerol
molecules, but excluding charged solutes, e.g., protons \cite{borgnia99-425}.
The recently determined 3D structure of GlpF at atomic resolution
\cite{fu00-481} (Fig.~\ref{fig:pmf}) has provided much
insight into the underlying microscopic mechanism of molecular transport
and selectivity through GlpF \cite{glpf1}.
In particular, molecular dynamics (MD) studies
\cite{groot01-2353,jensen01-1083,jensen02-6731,GRAY2003} established
that water and glycerol diffusion through GlpF is single file, and for
biologically relevant periplasmic glycerol concentrations correlation
effects between consecutive glycerol molecules are negligible due to
their large spatial separation.
The corresponding \emph{potential of mean force} (PMF)
\cite{jensen02-6731} that guides the transport of glycerol through the
channel is highly asymmetric reflecting the atomic structure of
GlpF (see Fig.~\ref{fig:pmf}), with a prominent potential well at the
external (periplasmic) side and a constriction region with several
pronounced potential barriers towards the internal (cytoplasmic) side of
the channel \cite{fu00-481,sansom01-R71}.
Besides GlpF, there are several other porins which exhibit similar spatial
asymmetry \cite{cowan92-727,wang97-56,forst98-37}, and in spite of recent
efforts, no biological function could be attributed to the asymmetry
\cite{LU2003}.
Here we propose and demonstrate that under realistic physiological
conditions, the asymmetry of GlpF furnishes active glycerol transport
through a \emph{ratchet-like} mechanism. In general, the ratchet effect
refers to the generation of directed motion of Brownian particles in a
spatially \emph{periodic} and \emph{asymmetric} (ratchet)
potential in the presence of \emph{non-equilibrium} fluctuations and/or
externally applied time-periodic force with zero mean
\cite{reimann02-57,magnasco93-1477,astumian97-917}.
Since cell membranes are subject to non-equilibrium fluctuations, which
span a wide range of time and length scales
\cite{levin91-733,pelling04-1147}, one expects a ratchet effect contribution
to the transport through asymmetric channel proteins, such as GlpF, even
if the PMF is non-periodic.
Assuming simplified, heuristic models, the ratchet effect has been
invoked before to explain the functioning of active biomolecules, e.g.,
motor proteins and ATP hydrolysis driven ion pumps
\cite{reimann02-57,tsong02-345}.
To our knowledge this Letter uses for the first time a realistic,
microscopically determined PMF to investigate the precise role of
non-equilibrium fluctuations in facilitated transport.
We find that as a result of channel asymmetry glycerol uptake, driven by a
concentration gradient, is enhanced significantly in the presence of
non-equilibrium fluctuations. Furthermore, the ratchet effect-caused
enhancement is larger for the outward, i.e., from the cytoplasm to the
periplasm, flux than for the inward one, suggesting that non-equilibrium
fluctuations also play an important role in protecting the interior of the
cell against excess uptake of glycerol.

Glycerol transport through GlpF can be modeled in terms of overdamped Brownian
motion along the axis of the channel as a result of the concentration gradient
established at the ends of the channel.  The interaction of a diffusing
glycerol molecule with the protein, solvent, lipid and other glycerol
molecules is taken into account through the PMF, $U(x)$, as determined from
steered molecular dynamics simulations by employing the Jarzynski equality
\cite{jensen02-6731}.
The motion of a glycerol molecule inside GlpF and in the presence of an
external force $F(t)$ in the strong friction limit is described by the
Langevin equation (LE): $\gamma\dot{x}=f(x)+\xi(t)+F(t)$,
where $\gamma$ is the friction coefficient, $f(x)=-U'(x)$ is the deterministic
force derived from the PMF, and $\xi(t)$ is the Langevin force due to the
equilibrium thermal fluctuations. As usual, $\xi(t)$ is a Gaussian white
noise with $\langle\xi(t)\rangle=0$ and $\langle\xi(t)\xi(0)\rangle =
2D\gamma^{2}\delta(t)$, where $\delta(t)$ is the Dirac-delta function, and $D$
is the effective diffusion coefficient of a glycerol molecule inside GlpF. According
to the \textit{fluctuation-dissipation} theorem, $D$ and $\gamma$ are related
through the Einstein relation $D=k_{B}T/\gamma$.
We assume that $F(t)$ is time-dependent, but homogeneous. It describes
either an externally applied force, or some intrinsic non-equilibrium
fluctuations of the system (see below). Due to the single file nature of
the glycerol transport through GlpF, a force applied at either end of
the channel will be transmitted along the file without significant loss
in intensity (\emph{incompressibility} of the single file), which
justifies our assumption that $F(t)$ is homogeneous along the channel.
For a periodic $f(x)$, %Eq.~(\ref{eq:1}) 
the above LE
describes what is often referred
to as a fluctuating force (tilting) ratchet \cite{reimann02-57}.

At this point we introduce dimensionless units that will be employed
throughout this paper, unless otherwise stated. All other units can be
expressed in terms of the following three: length of GlpF
${\mathcal{L}}=L\approx4.8$~nm, diffusion time ${\mathcal{T}} = \tau_{D}
= L^{2}/D\approx10^{-7}s$, and thermal energy
${\mathcal{E}}=k_{B}T\approx4.28\times10^{-21}~J$; here $k_{B}$ is the
Boltzmann constant, $T=310$~K is the physiological temperature, and
$D\approx2.2\times10^{-10}$ m$^{2}/$s is the effective diffusion
coefficient of glycerol inside GlpF \cite{LU2003}. Thus, the force unit
is ${\mathcal{F}}=\gamma D/L=k_{B}T/L\approx0.9~pN$.
In the new units, the Fokker-Planck equation (FPE) corresponding to
the above LE reads
\begin{subequations}
\begin{eqnarray}
\partial_{t}p(x,t)&=&-\partial_{x}J(x,t),
\label{eq:2a}\\
J(x,t)&=&-\partial_{x}p(x,t)+[f(x)+F(t)]p(x,t)
\label{eq:2b}
\end{eqnarray}
\label{eq:2}
\end{subequations}
where $p(x,t)dx$ is the (unnormalized) probability of finding a glycerol
molecule in $(x,x+dx)$ (see Fig \ref{fig:pmf}), and $J(x,t)$ is the local,
instantaneous flux of glycerol through the channel.
The probability density $p(x)$ is related to the local concentration
$C(x)$ by $p(x)=S(x)C(x)$, with $S(x)$ the area of the channel cross
section. 
From the crystal structure one finds that the opening area at both ends
of GlpF is $S_0 \equiv S(0) \approx S(1) \approx 100$ \AA$^{2}$
\cite{fu00-481}. 

\begin{figure}[!ht]
  \centering
  \includegraphics[width=3.2in]{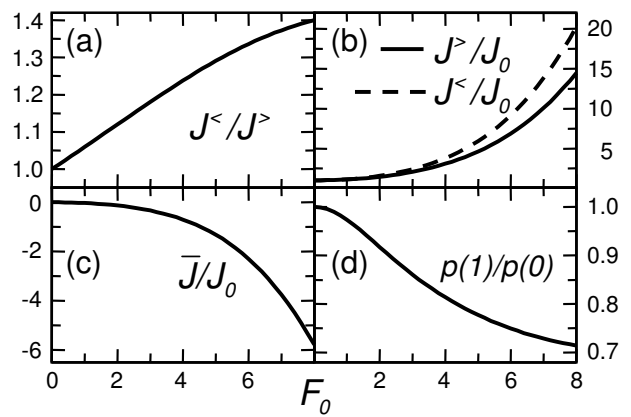}
  \caption{(a) Ratio of outward and inward glycerol fluxes in GlpF as function
    of a constant force $F_0$. (b,c) Transport induced by a square-wave
    force in GlpF. Shown are the inward [$p_1=0$], outward [$p_0=0$],
    and equal concentration [$p_1=p_0$] fluxes vs force amplitude $F_0$.
    (d) Ratio of the inner to outer glycerol concentrations vs $F_0$ for
    vanishing flux.}
  \label{fig:sw}
\end{figure}

GlpF can be regarded as a nanopore, which connects two reservoirs of
glycerol molecules located at $x=0$ (periplasm) and $x=1$ (cytoplasm),
respectively (Fig~\ref{fig:pmf}). The glycerol uptake is driven by a
concentration gradient: glycerol concentration (and therefore $p_0\equiv
p(0)$) is finite in the periplasm and vanishingly small in the cytoplasm
($p_1\equiv p(1)\approx 0$). Indeed, once glycerol enters the cytoplasm
it gets phosphorylated by glycerol kinase (GK) and, as a charged
particle, the product glycerol phosphate cannot leave the cell
\cite{voegele93-1087}. However, excessive accumulation of glycerol
phosphate may result in cell poisoning and death. To avoid this, the
enzyme GK is genetically turned off, preventing further glycerol
phosphorylation \cite{voegele93-1087}. Since the glycerol concentration
gradient across the cell membrane persists, one expects that glycerol
uptake should continue, in spite of its potential damaging effect on the
cell, until this gradient vanishes (i.e., $p_1=p_0$).  Below we
demonstrate that channel asymmetry, combined with non-equilibrium
fluctuations, can stop glycerol uptake against the persistent
concentration gradient, keeping the cytoplasmic glycerol at a level
$p_1<p_0$.
To this end, we calculate the steady glycerol flux through GlpF in four
distinct cases corresponding to suitable choices of $F(t)$.
\begin{widetext}

\begin{figure}[!th]
\includegraphics[width=7in]{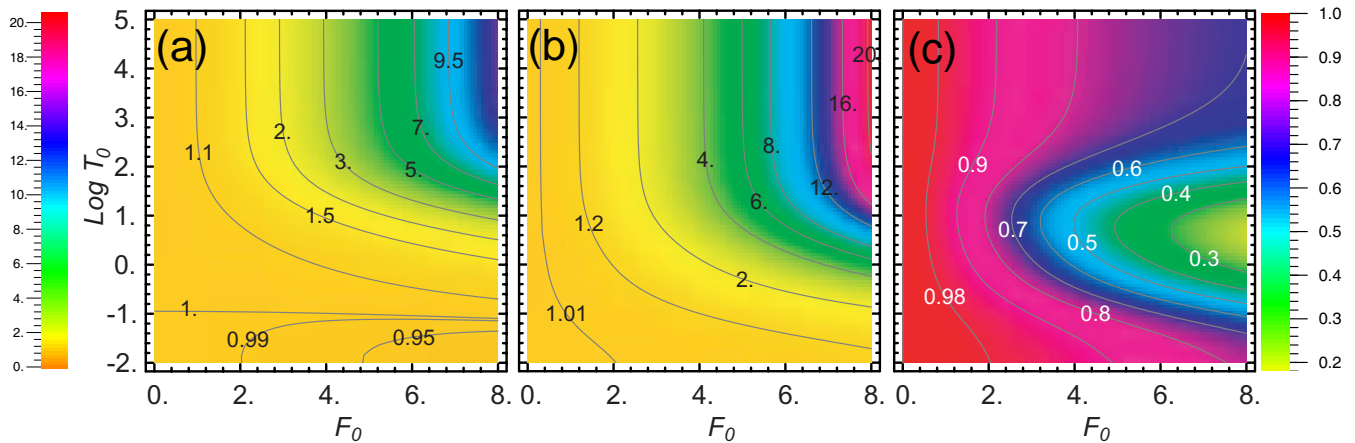}
\caption{(color online). Contour density plots of the numerically evaluated 
  (a) relative inward $J(F_0,T_0|p_0,0)/J_0$ and (b) outward
  $-J(F_0,T_0|0,p_0)/J_0$ glycerol fluxes in GlpF, and (c) concentration
  ratio $p_{1}/p_0={\cal A}_0/{\cal A}_1$ at which current reversal
  ($J=0$) takes place in the channel, as a function of the amplitude
  $F_0$ and mean switching time $T_0$ of the RTF. $J_0\equiv
  J(0,T_0|p_0,0)$ is the inward flux in the absence of the RTF. The
  scales on the left and right refer to (a)-(b) and (c), respectively.}
\label{fig:rtn-J}
\end{figure}

\end{widetext}

\emph{1) Transport driven by concentration gradient} [$F(t)=0$] -- In the
steady state, the flux is constant throughout the channel
[c.f.~Eq~\ref{eq:2a}] and, in the absence of the external force $F(t)$, is
given by
\begin{subequations}
\begin{eqnarray}
J(p_0,p_1) & = & A_{0}p_{0}-A_{1}p_{1},\label{eq:J0a}\\
A_{i} & = & e^{U_{i}}\left[\int_{0}^{1} e^{U(x)}\,dx\right]^{-1}, \quad
i=0,1\label{eq:J0b} 
\end{eqnarray}
\label{eq:J0}
\end{subequations}
Since the PMF $U(x)$ vanishes on both sides of the channel, i.e.,
$U_{0}\equiv U(0)=U(1)\equiv U_{1}=0$ (see Fig.~\ref{fig:pmf}), one has
$A_{0}=A_{1}$, and the flux is proportional to the glycerol
concentration difference, i.e.,
$J(p_0,p_1)=A_{0}\left(p_{0}-p_{1}\right)=A_{0}S_0\left(C_{0}-C_{1}\right)$.
Hence, the flux vanishes in the absence of a concentration gradient, and
the flux is insensitive to the asymmetry of the channel.
In order to produce directed transport, one needs to drive the system
out of equilibrium, e.g., by applying an external force or by subjecting
the system to non-equilibrium fluctuations.

\emph{2) Transport driven by potential gradient} [$F(t)=F_0=const.$] -- The
dependence of the flux on the asymmetry of the channel is manifest when
$U_0\neq U_1$, i.e., in the presence of a potential gradient. The constant
external force leads to an effective, tilted PMF, $U_{eff}(x) = U(x) - F_0 x$,
and according to Eqs.~\ref{eq:J0} the stationary flux reads
\begin{subequations}
\begin{equation}
\label{eq:Jca}
J\left(F_0|p_{0},p_{1}\right) =  {\cal A}_{0}(F_{0})\, p_{0}-{\cal A}_{1}(F_{0})\, p_{1}
\end{equation}
\begin{equation}
  \label{eq:Jcb}
  A_{0}(F_{0})  =   \left[\int_0^1
    e^{U_{eff}(x)} dx\right]^{-1},\quad
  A_{1}(F_{0}) =  A_0(F_0) e^{-F_0}
\end{equation}
\label{eq:Jc}
\end{subequations}
For a symmetric PMF, i.e., $U(1-x)=U(x)$, follows $A_1(-F_0)=A_0(F_0)$,
implying that the (magnitude of the) flux is also symmetric
$|J\left(F_0|p_0,p_1\right)| = |J\left(-F_0|p_1,p_0\right)|$.
In general, for a generic asymmetric PMF the inward and outward fluxes
will be different. Under normal conditions when $p_1=0$, the inward flux
is $J^>=A_0(F_0)p_0$. The same flux through an inverted channel, i.e.,
$(p_0,p_1) \rightarrow (0,p_0)$ and $F_0\rightarrow -F_0$ , would be
$J^<=A_1(-F_0)p_0$. It can be readily checked that $J^\gtrless / J_0>1$,
i.e., the constant driving force enhances the flux when applied along
the concentration gradient. Furthermore, according to
Fig.~\ref{fig:sw}a, under identical conditions the flux through the
inverted channel is always larger than the flux through the normally
oriented GlpF, the ratio of the two increasing monotonically with $F_0$.

\emph{3) Transport driven by an external periodic driving force} -- Next, we
consider an external force $F(t)$ that switches periodically between $\pm
F_0$, $F_0=const$ (\textit{square-wave} force).  Although the time average of
$F(t)$ is zero, this force induces a finite flux through GlpF even in the
absence of a concentration gradient. Indeed, assuming that one can neglect the
transient in the instantaneous flux after switching $F(t)$, the mean flux
$\bar{J}$ through the channel can be expressed as
$\bar{J}(F_0|p_0,p_1) = \left[J(F_0|p_0,p_1)+J(-F_0|p_0,p_1)\right]/2$,
where $J(F_0|p_0,p_1)$ is given by Eq.~\ref{eq:Jca}. Then, for
$p_1=p_0$, $\bar{J}/J_0\equiv \bar{J}(F_0|p_0,p_0)/\bar{J}(0|p_0,0)$ is
negative, and decreases monotonically with $F_0$, as shown in
Fig.~\ref{fig:sw}c. In this case too, $\bar{J}<0$ implies that for GlpF
the outward flux $J^<=-J(-F_0|0,p_0)$ is bigger than the inward flux
$J^>=J(F_0|p_0,0)$ as shown in Fig.~\ref{fig:sw}b, and both fluxes
$J^\gtrless$ are bigger than $J_0$, the flux in the absence of the
external force. Furthermore, the concentration ratio $p_1/p_0$ at which
the flux through the channel vanishes (current reversal), i.e.,
$\bar{J}(F_0|p_0,p_1)=0$, is plotted as a function of $F_0$ in
Fig.~\ref{fig:sw}d. The values $p_1/p_0<1$ found are expected and
consistent with the fact that for the same force level and concentration
gradient, the outward flux is larger than the inward flux.

\emph{4) Transport driven by non-equilibrium fluctuations} -- Finally,
we consider the effect of non-equilibrium fluctuations of the cell
membrane on the glycerol transport through GlpF.
We model such fluctuations by a \emph{random telegraph force} (RTF),
i.e., a homogeneous \textit{dichotomous} force $F(t)$, which switches
between two states $\pm F_{0}$ with switching times that obey a Poisson
distribution. For the RTF holds $\langle F(t)\rangle=0$, and $\langle
F(t)F(0)\rangle=F_{0}^{2}e^{-2t/T_0}$, where $T_0$ is the mean switching
time.  
The stationary FPE in this case consists of two coupled equations,
$-\partial_x^2 p_{\pm}(x)+\partial_x [f(x)p_{\pm}(x)] \pm F_0 p_{\pm}(x)
-p_{\pm}(x)/T_0 + p_{\mp}(x)/T_0=0$, where $p_{\pm}(x)$ is the
conditional probability density that the RTF is in the $\pm F_0$ state.
The corresponding flux is $J=-p'(x)+f(x)p(x)+F_0\Delta p(x)=const$,
where $p=p_+ + p_-$ is the total probability density, and $\Delta p
= p_+ - p_-$.
For the feature-rich GlpF PMF (Fig.~\ref{fig:pmf}) the flux
$J(F_0,T_0|p_0,p_1)={\cal A}_0(F_0,T_0)p_0-{\cal A}_1(F_0,T_0)p_1$ needs
to be computed numerically, e.g., by employing a matrix
continued-fraction method.
The computed flux is shown in Fig.~\ref{fig:rtn-J} as two dimensional
density plots for $F_0\in[0,8]$ and $T_0\in[10^{-2},10^5]$ (logarithmic
scale). In SI units, these values correspond to $F_0\in[0,7.2]$~pN and
$T_0\in[10^{-9},10^{-2}]$~s, respectively. 
The conclusions drawn from Fig.~\ref{fig:rtn-J} are consistent with our
previous findings for externally applied deterministic forces. First,
the RTF-induced asymmetry between the inward and outward fluxes is
manifest (Figs.~\ref{fig:rtn-J}a-b), with bigger outward flux for
the same $F_0$, $T_0$ and concentration gradient. Second, the
RTF-induced flux enhancement is more pronounced for slower fluctuations
and for larger $F_0$.  Third, the flux in the absence of a concentration
gradient across the membrane is always outward. Thus, the ratio of the
inner to outer glycerol concentrations at which glycerol uptake ceases
is always less than one, as shown in Fig.~\ref{fig:rtn-J}c.

Our calculations have demonstrated that non-equilibrium force
fluctuations acting on glycerol in GlpF can have an important effect on
the glycerol uptake by a cell. On the one hand, slow, large amplitude
fluctuations enhance the concentration gradient-driven glycerol uptake
(Fig.~\ref{fig:rtn-J}a), which may be vital for the cell under poor nutrient
conditions. On the other hand, when glycerol is abundant, fluctuations
provide an effective protection mechanism to the cell by stopping
glycerol uptake well before the cytoplasmic concentration reaches the
periplasmic one (Fig.~\ref{fig:rtn-J}c).
The mechanism underlying this behavior is related to the ratchet effect
and depends sensitively on the asymmetric shape of the PMF
characterizing glycerol conduction in GlpF. 
The effects described should be testable experimentally. In fact, recent
experiments have demonstrated that cell membranes are subject to slow,
kHz frequency ($T_0\sim 10^{-2}$) non-equilibrium fluctuations that may
be related to pumping mechanisms by which cells supplement the passive
diffusion of nutrients \cite{pelling04-1147}.

The authors thank Emad Tajkhorshid for valuable insights.  We
acknowledge financial support from the Univ.~of Missouri Research Board,
the Institute for Theoretical Sciences and the Department of Energy (DOE
Contract W-7405-ENG-36) for I.K., and the National Institutes of Health
(NIH 1-R01-GM067887-01) for K.S.

%---------------------

\end{document}